\documentstyle[aps,prb,twocolumn]{revtex}
\draft
\begin{document}
\def\vec#1{{\bf{#1}}}
\title{Algebraic approach to renormalization}
\author{Jochen Rau\thanks{Address after October 1, 1996:
Max-Planck-Institut f\"ur Physik komplexer Systeme,
Bayreuther Stra{\ss}e 40 Haus 16,
01187 Dresden, Germany.}}
\address{European Centre for Theoretical Studies
in Nuclear Physics and Related Areas (ECT$^*$), \\
Villa Tambosi,
Strada delle Tabarelle 286, 38050 Villazzano (Trento), Italy}
\date{July 27, 1996}
\maketitle
\begin{abstract}
In close analogy to the Bloch-Feshbach formalism known from
the theory of nuclear dynamics, I develop a mathematical framework
that allows one to understand renormalization 
in terms of purely algebraic operations (projections, dilatations)
in Hilbert space. 
This algebraic approach is put to the test in the study of
the low-energy dynamics of interacting quantum gases,
and proves to be efficient in
deriving such diverse results as the renormalization group equation
for an interacting
Bose gas, the $\beta$ function of $\phi^4$ theory, the
screening of fermion-fermion interactions or the BCS instability.
\end{abstract}
\medskip
\narrowtext
\section{Introduction}
The idea of
eliminating irrelevant modes in order to focus on the
dynamics of few selected degrees of freedom has a long history.
The Bloch-Feshbach formalism, developed in the late 50's to
describe selected features of nuclear dynamics,\cite{bloch,haw}
employs projection operators in Hilbert space
in order to determine the effective Hamiltonian in some
restricted model space, thereby discarding dynamical information
that pertains to the irrelevant modes.
The irrelevant modes no longer appear explicitly
in the effective theory, yet their residual influence on the
dynamics of the remaining modes is taken into account through
adjustments of the effective interaction.
 
In a similar spirit, 
Wilson's renormalization group, developed in the early 70's
to better understand critical phenomena,\cite{kadanoff,wilson,wegner,polchinski}
is a mathematical tool that allows one to 
iteratively eliminate short-wavelength modes and thus to arrive at
effective (``renormalized'')
theories which describe the dynamics on successively
larger length scales.
In the original context of second-order phase transitions the
renormalization group mainly served 
to calculate critical exponents, and
to provide a satisfactory
theoretical explanation for their universality.
More recently, it has been pointed out that
the renormalization group rationale also affords a natural framework
for the physics of interacting fermion systems, and that it helps to
understand such diverse phenomena as Landau's fermi liquid theory,
charge-density waves, BCS instability, or 
screening.\cite{polchinski_tasi,weinberg,shankar,dupuis}

The two calculational tools for systematic mode
elimination 
--on the one hand
the Bloch-Feshbach
projection technique in Hilbert space,
based on algebraic concepts such as
linear subspaces and operators,
and on the other hand Wilson's renormalization group,
commonly formulated in terms of 
functional integrals and Feynman diagrams--
both derive their power 
from the fact that they allow one to study 
accurately selected features
of the dynamics (for instance its infrared limit)
without ever having to solve the
full underlying microscopic theory.
Beyond this common root for their success, mounting evidence like
the success of Anderson's 
``poor man's scaling'' approach to the Kondo problem,\cite{anderson}
Seke's projection-method treatment of the nonrelativistic
Lamb shift,\cite{seke}
recent studies of a simple quantum
mechanical model,\cite{fields}
or a calculation of the 1-loop renormalization of $\phi^4$ theory
with the help of Bloch-Feshbach techniques,\cite{muller:rau}
suggests that the two methods are very closely
related, and that in some cases the latter (Wilson) can even be regarded as
a special case of the former (Bloch-Feshbach).

It is the purpose of the present paper
to propose a slightly modified version of the Bloch-Feshbach formalism
as an efficient tool to do renormalization (Sec. \ref{proj_technique}).
In the literature it has been put forth that the practical use of such a
projection operator approach
was limited to simple, essentially
one-dimensional problems,
and that the study of more complicated physical systems
like a Fermi liquid
would instead require the use of
functional integral methods.\cite{shankar}
Contrary to this assertion, I wish to show how
algebraic techniques in Hilbert space can be successfully
employed to tackle a variety of non-trivial phenomena
in more than one dimension.
I shall focus on the application to interacting quantum gases
(Sec. \ref{gases}), both Bose and Fermi,
and illustrate the versatility of the algebraic approach
in deriving four well-known results:
the flow of the coupling constant for
bosons with point interaction;
closely related, the one-loop $\beta$ function of
$\phi^4$ theory;
the screening of fermion-fermion interactions;
and the BCS instability (Sec. \ref{examples}).

Despite the obvious parallels between the Bloch-Feshbach formalism
and Wilson's renormalization,
both in spirit and --as I will show-- in practical calculations,
there remains however an important difference:
beyond the elimination of
short-wavelength modes,
which is common to both approaches,
the renormalization program generally requires 
the additional step of rescaling.
How this additional step, too, can be implemented within the
algebraic framework, and under which circumstances it is called for,
will be discussed in a separate section
(Sec. \ref{renormalization}) at the end of the paper.
\section{Projection Technique}\label{proj_technique}
We consider a macroscopic system (e. g., an interacting
quantum gas) whose possible microstates span a
--typically large-- Hilbert space (e. g., boson
or fermion Fock space),
and whose dynamics in this Hilbert space is governed
by some Hamiltonian $H$.
In equilibrium 
the state of the system is
described by a canonical statistical operator
\begin{equation}
\rho(b)= \exp[-H/b-\ln Z(b)]
\quad,\quad
b\ge 0
\label{canonical}
\end{equation}
with partition function $Z(b)$,
where $b$ may represent the temperature $T$, Boltzmann's constant $k_B$, 
their product $k_B T$,
or possibly some other
quantity which we choose to keep explicitly as an external parameter;
it will be specified later.
All other parameters, such as masses, coupling constants,
the chemical potential or --in case $b\not\in \{T,k_B T\}$-- the temperature, are absorbed
into the definition of the Hamiltonian.
Given the statistical operator as a function of $b$,
the underlying Hamiltonian can be extracted via
\begin{equation}
b^2 {\partial\over\partial b}\rho(b)=
[H+c(b)] \rho(b)
\quad.
\label{hamiltonian}
\end{equation}
Here $c(b)$ is a $c$-number which stems from the
derivative of the partition function.
Its specific value need not concern us,
as long as we disregard simple shifts of the ground state energy.
The partial derivative with respect to the external parameter is taken at
fixed masses, couplings, chemical potential, and other parameters
in the Hamiltonian.
 
We presume that we are only interested in certain selected
features of the macroscopic system (e. g., its long-wavelength
properties), and that these selected features are represented
by observables which merely act in some subspace of the original
Hilbert space (e. g., in the subspace spanned by many-particle states
with all momenta below some cutoff).
Let the operator which projects the 
original Hilbert space
onto this selected subspace be denoted by $P$, and 
its complement by $Q=1-P$.
Information pertaining to the selected degrees of freedom
is then entirely encoded in the
reduced statistical operator\cite{trace}
\begin{equation}
\rho_{\rm eff}(b):=P\rho(b) P
\quad;
\end{equation}
the other parts of the original density matrix,
$P\rho Q$, $Q\rho P$ and $Q\rho Q$, only carry information which
to us is irrelevant.\cite{info_loss}
 
We would now like to cast the reduced statistical operator again
into the canonical form (\ref{canonical}), with some
modified, effective Hamiltonian $H_{\rm eff}$ which acts
in the smaller subspace only ($H_{\rm eff}\equiv PH_{\rm eff}P$).
This effective Hamiltonian will in general acquire a
dependence on the parameter $b$, so we must write
\begin{equation}
\rho_{\rm eff}(b)=
P\,\exp[-H_{\rm eff}(b)/b -\ln Z_{\rm eff}(b)]
\quad.
\end{equation}
We assume, however, that the effective Hamiltonian is a smooth function
of $b$ which can be Taylor expanded around $b=0$.
For small parameter values we may then set,
as a first approximation,\cite{loops}
\begin{equation}
H_{\rm eff}(b)\approx H_{\rm eff}(0)
\quad.
\end{equation}
This approximate effective Hamiltonian can be determined
in analogy to Eq. (\ref{hamiltonian}), via
\begin{equation}
\left.b^2 {\partial\over\partial b}\rho_{\rm eff}(b)
\right|_{b=0}=
[H_{\rm eff}(0)+c'] \rho_{\rm eff}(0)
\quad,
\label{eff_hamiltonian}
\end{equation}
again up to some (generally different) $c$-number.

In order to evaluate the left-hand side of the equation
(\ref{eff_hamiltonian}), we first use (\ref{hamiltonian}) to obtain
\begin{equation}
b^2 {\partial\over\partial b}\rho_{\rm eff}(b)
=
[PHP+c]\rho_{\rm eff}(b) + PHQ\rho(b)P
\label{start}
\end{equation}
which includes a new term $PHQ\rho(b)P$ that accounts for the
overlap between the $P$- and $Q$-sectors.
Assuming that the original Hamiltonian can be decomposed into a free
part and an interaction part,
\begin{equation}
H=H^{(0)}+V
\quad,
\end{equation}
where the free part $H^{(0)}$ commutes with the
projection, 
this overlap term is $O(V^2)$ and hence expected to be small
for a weak interaction.
With the help of the identity
\begin{eqnarray}
Q\rho(b)P
&=&
-Q\int_0^{1/b} d\lambda\,{d\over d\lambda}
\left[e^{-\lambda H}Qe^{\lambda H}\right] \rho(b) P
\nonumber \\
&=&
-Q\int_0^{1/b} d\lambda\, e^{-\lambda H} [Q,V] e^{\lambda H} \rho(b) P
\end{eqnarray}
we find,
to lowest nontrivial order in perturbation theory,
\begin{equation}
Q\rho(b)P
=
Q\left[ {\exp(-{\cal L}^{(0)}/b)-1 \over {\cal L}^{(0)}}(QVP) \right]
\rho_{\rm eff}(b)
\quad.
\label{intermed}
\end{equation}
Here ${\cal L}^{(0)}$ denotes 
the Liouville ``super''operator which, when acting on an arbitrary
Hilbert space operator $A$, takes the
commutator with the free Hamiltonian:
\begin{equation}
{\cal L}^{(0)} A := [H^{(0)},A]
\quad.
\end{equation}
Since
\begin{equation}
{\cal L}^{(0)}|E^{(0)}_i\rangle \langle E^{(0)}_j| =
(E^{(0)}_i-E^{(0)}_j) |E^{(0)}_i\rangle \langle E^{(0)}_j| 
\end{equation}
for free energy eigenstates $\{|E^{(0)}_i\rangle\}$, it
may be viewed as probing the energy difference
between in- and out-states.
 
Provided the states in the $Q$-sector have higher energies than those
in the $P$-sector, the equation (\ref{intermed}) has a well-defined
limit $b\to 0$: simply, $\exp(-{\cal L}^{(0)}/b)\to 0$.
Inserting this limit into Eq. (\ref{start}) then yields
\begin{equation}
H_{\rm eff}(0)=
PHP+ \Sigma + c''
\quad,
\label{eff_h}
\end{equation}
with $\Sigma$ given to second order perturbation theory by
\begin{equation}
\label{approx_eff}
\Sigma^{(2)} = -PVQ{1\over {\cal L}^{(0)}}QVP
\quad.
\end{equation}
Clearly, the effective Hamiltonian is not just the
projection $PHP$ of the original Hamiltonian, but contains
also the extra term $\Sigma$
to account for the residual influence of the eliminated $Q$-modes.
\section{Application to Interacting Quantum Gases}\label{gases}
\subsection{Mode Elimination} 
We are interested in the low-temperature properties 
of interacting quantum gases, and hence in the effective
dynamics of low-energy excitations above
the many-particle ground state.
For noninteracting bosons
the ground state has all particles
in the lowest energy, zero momentum single-particle mode; 
while for noninteracting fermions the ground state 
consists of a filled Fermi sea,
with all momentum modes occupied up to some 
Fermi momentum $K_F$.
(For simplicity, the Fermi surface will always be taken to be spherical.)
We will assume that, at least to a good approximation, 
these essential features of the ground state survive
even in the presence of interaction.
More specifically, we will assume that in the case of interacting
bosons the ground state still has most particles in modes
with zero, or at least very small, momentum;
and that in the case of interacting fermions 
there still exists a well-defined 
Fermi surface.
Low-energy excitations then correspond to the promotion of bosons 
from small to some slightly higher momentum,
or of fermions from just below the Fermi surface to just above it.\cite{landau}
At low temperature the regions of interest in momentum space are therefore
the vicinity of the origin (bosons) or the vicinity of the Fermi 
surface (fermions), respectively.

We now wish to devise a systematic procedure for focusing onto 
these regions of interest.
To this end we consider effective theories 
(i) in the bosonic case for modes within a
sphere around the origin, of radius $\Lambda$; and
(ii) in the fermionic case for modes within a shell 
inclosing the Fermi surface, of mean radius $K_F$ and
thickness $2\Lambda$ (where $\Lambda\ll K_F$).
Whereas in the limit of large $\Lambda$ one recovers the original,
full theory, the opposite limit $\Lambda\to 0$ 
yields the desired low-energy effective theory.
In order to interpolate between these two limits
we proceed in infinitesimal steps:
we lower the 
cutoff from some given $\Lambda(s)$ to
\begin{equation}
\Lambda(s+\Delta s):=\exp(-\Delta s)\Lambda(s)
\quad,\quad \Delta s\ge 0
\label{cutoff_reduction}
\end{equation}
with $\Delta s$ infinitesimal,
thereby discarding from the theory momentum modes pertaining
to an infinitesimal shell (in the fermionic case: two shells) 
of thickness $\Delta\Lambda=\Lambda(s)\,\Delta s$;
determine the effective dynamics of the remaining modes;
then eliminate the next infinitesimal shell, 
again determine the effective dynamics of the remaining modes,
and so on.
After each infinitesimal step we obtain a new 
effective Hamiltonian with slightly modified coupling constants.
These may also include novel couplings which had not been present
in the original theory; in fact, the mode elimination
procedure will typically generate an infinite number of such
novel couplings.
But in many cases only a few coupling constants will change
appreciably and thus suffice to study the physical system at hand.
How these coupling constants evolve as the flow parameter
$s$ increases and hence the cutoff $\Lambda(s)$ approaches zero,
can then be described by a small set of coupled
differential equations. 
Modulo rescaling, which I shall discuss separately in Sec.
\ref{renormalization}, these are the
{\em renormalization group equations} of the theory.

Whether or not the temperature is among the quantities that
flow as a function of $s$, depends on whether it has been included
in the definition of the external parameter $b$. 
If $b$ is chosen to be $k_B$ or some other
temperature-independent quantity then the temperature may flow,
like all variables which have been absorbed into the
definition of the Hamiltonian.
If $b=T$ or $b=k_B T$, on the other hand, then the temperature is
regarded as a parameter that is controlled externally and
hence fixed: it does not change upon mode elimination.
While the former scenario 
applies to the study of isolated systems which,
depending on the modes selected, may exhibit effective dynamics at
varying apparent temperatures,
the latter scenario 
is adapted to the study of systems which are
coupled to a heat bath of prescribed temperature.
It is this latter case which we shall consider,
as we wish to study the effective
dynamics in the low-temperature limit $T\to 0$ and thus explicitly control 
the temperature. Consequently, we choose $b=k_B T$. 

Our aim is now to illustrate in a few prominent cases
--bosons with point interaction, $\phi^4$ theory, screening of fermion-fermion
interactions, BCS instability--
how the applicable renormalization group equations
can be derived efficiently with the help of our
projection technique.
To this end we must specify the appropriate many-particle
Hilbert space (Fock space) and the appropriate projection operator
for each infinitesimal elimination step.
 
At a given cutoff $\Lambda$ the boson Fock space is spanned
by the particle-free vacuum $|0_{\rm b}\rangle$ and all $n$-particle states
($n=1\ldots\infty$)
\begin{equation}
|\vec{k}_1\ldots \vec{k}_n\rangle \propto
\prod_{i=1}^n a^\dagger(\vec{k}_i) |0_{\rm b}\rangle
\quad,\quad
|\vec{k}_i|\le \Lambda
\quad,
\label{bose_excite}
\end{equation}
where the $\{{\vec{k}}_i\}$ denote the particle 
momenta and $\{a^\dagger(\vec{k}_i)\}$ the associated bosonic creation
operators.

The fermion Fock space, on the other hand, is spanned by
the filled Fermi sea (fermionic vacuum) $|0_{\rm f}\rangle$
and all its excitations which have particles above the Fermi surface
and/or missing particles (``holes'') below it,
all within a shell of thickness $2\Lambda$.
In order to cast this into a mathematical formulation
it is convenient to change coordinates,
from the true particle momenta $\{\vec{K}_i\}$ to little 
(``quasiparticle'') momenta
\begin{equation}
\vec{k}_i:=(|\vec{K}_i|-K_F)\,\hat{\vec{K}_i}
\end{equation}
and additional discrete labels
\begin{equation}
\sigma_i:=\mbox{sign}(|\vec{K}_i|-K_F)
\quad.
\end{equation}
This coordinate transformation $\vec{K}\to (\vec{k},\sigma)$
is invertible except for modes which
lie exactly on the Fermi surface.
States above the Fermi surface are labeled 
$\sigma=1$, while those below are labeled $\sigma=-1$.
The allowed excitations in fermion Fock space then
have the form ($n=1\ldots\infty$)
\begin{eqnarray}
|\vec{k}^\pm_1\ldots \vec{k}^\pm_n\rangle 
&\propto&
\prod_{i=1}^n [\theta(\sigma_i)a^\dagger(\vec{k}_i,\sigma_i)
+\theta(-\sigma_i)a(-\vec{k}_i,\sigma_i)] |0_{\rm f}\rangle
\, ,
\nonumber \\
&&\quad \quad \quad \quad \quad
|\vec{k}^\pm_i|\le \Lambda
\quad,
\label{fermi_excite}
\end{eqnarray}
where the $\{{\vec{k}}^\pm_i\}$ denote the  
momenta of particles ($+$) or holes ($-$), respectively,
and $\{a^\dagger\}$, $\{a\}$ the associated fermionic creation
and annihilation operators.
For simplicity,
we have omitted any spin quantum numbers.

It is now obvious which form the projection operator will have
that is associated with the infinitesimal cutoff reduction
(\ref{cutoff_reduction}): if applied to any of the excitations
(\ref{bose_excite}) or (\ref{fermi_excite}) 
it will simply multiply the respective state by a product of $\theta$ functions,
$\prod_i \theta(\Lambda-e^{\Delta s}|\vec{k}^{(\pm)}_i|)$,
to enforce the new cutoff constraint.
\subsection{Hamiltonian}
We consider interacting quantum gases whose dynamics in the
original, full Hilbert space is governed by a Hamiltonian
of the form
\begin{eqnarray}
H&=&\sum_k \epsilon_k :\!a^\dagger_k a_k\!:
+{1\over4}\sum_{ijkl}
\langle lk|V|ji\rangle_\pm :\!a^\dagger_l a^\dagger_k a_j a_i\!:
\nonumber \\
&=&H^{(0)}+V
\quad,
\label{quantum_gas}
\end{eqnarray}
with kinetic energy $H^{(0)}$
and a two-body interaction $V$.
By definition the single-particle energies $\epsilon_k$ include
the chemical potential.
The annihilation and creation operators obey
\begin{equation}
[a_i,a^\dagger_j]_\mp = \delta_{ij}
\label{commutation}
\end{equation}
for bosons (upper sign) or fermions (lower sign),
respectively.
Each term in the Hamiltonian is normal ordered
($:\ldots:$) with respect to the many-particle ground state.
In the bosonic case this just coincides with the usual normal ordering:
all annihilation operators to the right, all creation
operators to the left. The explicit normal ordering of the
Hamiltonian is then redundant.
In the fermionic case, on the other hand, normal ordering means
shuffling all operators which annihilate the fermionic vacuum
($a_i$ with $\sigma_i=1$ or $a^\dagger_j$ with $\sigma_j=-1$) to the
right, all others
($a^\dagger_k$ with $\sigma_k=1$ or $a_l$ with $\sigma_l=-1$) to the left,
thereby changing sign depending on the degree of the permutation.
The explicit normal ordering of the Hamiltonian then becomes nontrivial.
In both cases the normal ordering ensures that the energy
of the (bosonic or fermionic) vacuum is set to zero,
\begin{equation}
\langle 0_{\rm b,f}|H|0_{\rm b,f}\rangle =0
\quad.
\end{equation}
\subsection{Modification of the 2-Body Interaction}
As we discussed earlier, each mode elimination will yield an
effective Hamiltonian that will generally contain a slightly altered mass,
chemical potential, two-body interaction, etc., and possibly
new interactions such as an effective three-body interaction.
Here we shall restrict our attention to the
modification of the two-body interaction.
In order to determine this modification
we must consider the general formula (\ref{eff_h})
for the effective Hamiltonian. As we have seen, 
the effective Hamiltonian is the sum of the projected
original Hamiltonian, $PHP$, and an extra term $\Sigma$ that accounts
for the residual influence of the eliminated modes.
For our purposes
the projection $PHP$ of the original Hamiltonian need not concern
us, as it will not lead to any modification of the two-body
interaction. 
Rather, we must investigate the ramifications of the
extra term $\Sigma$.

Application of the perturbative result (\ref{approx_eff}) yields:
\begin{eqnarray}
\Sigma^{(2)}
&=&
-{1\over16}\sum_{abcd} \sum_{ijkl}
\langle lk|V|ji\rangle_\pm \langle dc|V|ba\rangle_\pm 
\nonumber \\
&& \times
P:\!a^\dagger_l a^\dagger_k a_j a_i\!: Q{1\over {\cal L}^{(0)}} 
Q:\!a^\dagger_d a^\dagger_c a_b a_a\!: P
\quad.
\label{full_sigma}
\end{eqnarray}
The two projectors $P$ at both ends of the operator product
ensure that all external momenta lie below
the new, reduced cutoff;
whereas the projectors $Q$ in the center force at least one internal momentum
to lie in the infinitesimal shell which has just been eliminated.
Therefore, at least one pair of field operators must pertain
to the eliminated $Q$-modes, and hence be contracted:
$a^{(\dagger)}a^{(\dagger)}\to 
\langle 0_{\rm b,f}|a^{(\dagger)}a^{(\dagger)}|0_{\rm b,f}\rangle$.
The product of the remaining six field operators can then be rearranged
with the help of Wick's theorem, to yield a
decomposition
\begin{equation}
\Sigma^{(2)}=\Sigma^{(2)}_6+\Sigma^{(2)}_4
+\Sigma^{(2)}_2+\Sigma^{(2)}_0
\end{equation}
where each $\Sigma^{(2)}_n$ is normal ordered and contains
$n$ field operators.
The various terms shift the ground state energy ($n=0$),
modify the mass, the chemical potential, or more generally the form of the
single-particle dispersion relation ($n=2$), modify the
two-body interaction ($n=4$), and generate a new effective
three-body interaction ($n=6$).

Since we want to focus on the modification of the two-body interaction
we consider only the term with $n=4$.
Calculating this term from Eq. (\ref{full_sigma}) 
involves two contractions, at least one
of which must pertain to the eliminated $Q$-modes (see above),
and both of which must go across the central $Q(1/{\cal L}^{(0)})Q$
(due to momentum conservation).
Neglecting the energy of the external modes,
we find for bosons
\begin{equation}
\Delta \langle lk|V|ji\rangle_+
= -\Delta \left[\sum_{ab} {1\over 2(\epsilon_a+\epsilon_b)}
\langle lk|V|ba\rangle_+ \langle ab|V|ji\rangle_+\right]
\label{mod_bosons}
\end{equation}
and for fermions
\begin{equation}
\Delta \langle lk|V|ji\rangle_-
= \Delta^{(\rm ZS)}_{lk|ji} + \Delta^{(\rm ZS')}_{lk|ji} 
+ \Delta^{(\rm BCS)}_{lk|ji}
\label{fermion_mod}
\end{equation}
with
\begin{eqnarray}
\Delta^{(\rm ZS)}_{lk|ji}
&=& -\Delta [\sum_{ab}
{\theta(\sigma_a)\theta(-\sigma_b)-\theta(-\sigma_a)\theta(\sigma_b)
\over \epsilon_a-\epsilon_b}
\nonumber \\
&&
\quad
\times
\langle la|V|bi\rangle_- \langle bk|V|ja\rangle_-
]
\quad,
\label{zs}
\end{eqnarray}
its cross term
\begin{equation}
\Delta^{(\rm ZS')}_{lk|ji} = - \Delta^{(\rm ZS)}_{kl|ji}
\label{zs_prime}
\end{equation}
and
\begin{eqnarray}
\Delta^{(\rm BCS)}_{lk|ji}
&=& 
-\Delta [\sum_{ab}
{\theta(\sigma_a)\theta(\sigma_b)
- \theta(-\sigma_a)
\theta(-\sigma_b)
\over 2(\epsilon_a+\epsilon_b)}
\nonumber \\
&&
\quad
\times
\langle lk|V|ba\rangle_- \langle ab|V|ji\rangle_-
]
\quad.
\label{bcs}
\end{eqnarray}
The $\Delta$ in front of the sums signifies that
at least one of the internal modes ($a,b$) must 
lie in the eliminated shell.
In the bosonic case the modification of the two-body
interaction can be associated with a 1-loop ``ladder'' diagram;
in the fermionic case, on the other hand, there are three distinct contributions
which, with hindsight, can be identified with 
``zero sound'' (ZS,ZS') and BCS diagrams.\cite{shankar}
The ZS contribution and its cross term ZS' account for particle-hole
excitations ($\sigma_a=\pm 1,\sigma_b=\mp 1$),
while the BCS term describes 2-particle
($\sigma_a=\sigma_b=+1$) or 2-hole ($\sigma_a=\sigma_b=-1$)
excitations.
\section{Examples}\label{examples}
\subsection{Bosons with point interaction}
For spinless bosons with point interaction
($\delta$ function potential in real space) it is
\begin{equation}
\langle lk|V|ji\rangle_+
={2U\over\Omega}\, \delta_{\vec{k}_i+\vec{k}_j,\vec{k}_k+\vec{k}_l}
\quad,
\label{point_coupling}
\end{equation}
with the Kronecker symbol enforcing momentum conservation,
$\Omega$ being the spatial volume,
and $U$ the coupling constant.
Provided the magnitude of the external momenta $\vec{k}_i$, $\vec{k}_j$ 
is negligible compared to
the cutoff $\Lambda$, 
momentum conservation implies that
the internal modes $a,b$ must {\em both} lie in the
eliminated shell, and that hence 
$\epsilon_a=\epsilon_b=\epsilon_\Lambda$.
Application of the general formula (\ref{mod_bosons}) then yields
\begin{equation}
\Delta U= -{U^2 \over 2\Omega\epsilon_\Lambda}
\Delta \left[
\sum_{|\vec{k}_a|,|\vec{k}_b|\le\Lambda}
\delta_{\vec{k}_i+\vec{k}_j,\vec{k}_a+\vec{k}_b}
\right]
\quad,
\label{point_mod}
\end{equation}
where the sum
\begin{eqnarray}
\Delta \left[
\sum_{|\vec{k}_a|,|\vec{k}_b|\le\Lambda}
\delta_{\vec{k}_i+\vec{k}_j,\vec{k}_a+\vec{k}_b}
\right]
&\approx& 
\Delta\left[\sum_{|\vec{k}_a|,|\vec{k}_b|\le\Lambda}
\delta_{\vec{k}_b,-\vec{k}_a}\right]
\nonumber \\
&=& 
\sum_{|\vec{k}_a|\in [\Lambda-\Delta\Lambda,\Lambda]} 1
\end{eqnarray}
simply counts the number of eliminated states.
For a spherical cut in momentum space this number
of states is given by
\begin{equation}
\sum_{|\vec{k}_a|\in [\Lambda-\Delta\Lambda,\Lambda]} 1 =
\rho(\epsilon_\Lambda)\,{d\epsilon_\Lambda\over d\Lambda}\,
\Delta\Lambda
\quad,
\label{number_states}
\end{equation}
with $\rho(\epsilon_\Lambda)$ denoting the density of states at
the cutoff.
With $\Delta\Lambda=\Lambda\,\Delta s$ we thus obtain the
flow equation
\begin{equation}
\Delta U= 
-{d\ln\epsilon_\Lambda \over d\ln\Lambda}\,
{\rho(\epsilon_\Lambda)\over 2\Omega}\,
U^2\cdot \Delta s
\quad.
\label{boson_flow}
\end{equation}
 
For a dilute gas of nonrelativistic bosons 
in three spatial dimensions, with mass $m$,
dispersion relation $\epsilon_\Lambda=\Lambda^2/2m$
and density of states $\rho(\epsilon_\Lambda)=\Omega m\Lambda/2\pi^2$
the flow equation reduces to
\begin{equation}
\Delta U= - {m\Lambda \over 2\pi^2} U^2\cdot \Delta s
\quad.
\label{u_flow}
\end{equation}
By its very definition the sequence of effective theories retains
complete information about the system's low-energy dynamics.
Observables pertaining to this low-energy
dynamics 
are therefore unaffected by the successive mode elimination, and
hence independent of $s$.
For example, the scattering length\cite{abrikosov}
\begin{equation}
{a} = {m\over 4\pi}\left[ U(s) - U(s)^2 \int_{|\vec{p}|\le\Lambda(s)} 
{d^3p\over(2\pi)^3}\, {m\over \vec{p}^2} \right]
\end{equation}
stays constant under the flow (\ref{u_flow}), as
the $s$-dependence of the parameters $U$ and $\Lambda$
just cancels out
(up to third order corrections).
\subsection{The link to $\phi^4$ theory} 
There is an interesting relationship between the result 
(\ref{boson_flow})
and the 1-loop $\beta$ function for real $\phi^4$ theory.
The $\phi^4$ Hamiltonian describes the dynamics of coupled anharmonic
oscillators. It reads, in three spatial dimensions,
$H=H^{(0)} + V$
with kinetic energy
\begin{eqnarray}
H^{(0)}
&=&
{1\over2}\int d^{3}x :\!\left[\pi(x)^2
+ |\nabla\phi(x)|^2 + m^2\phi(x)^2\right]\!:
\nonumber \\
&=&
\sum_{\vec k} \epsilon_{\vec k}\, a^\dagger_{\vec k} a_{\vec k}
\end{eqnarray}
and interaction
\begin{eqnarray}
V
&=&
{g\over 4!} \int d^{3}x\,\phi(x)^4
\nonumber \\
&=&
{g\over 4!\Omega}
\sum_{\vec{k}_1\vec{k}_2\vec{k}_3\vec{k}_4}
\prod_{\alpha=1}^4
{1\over\sqrt{2\epsilon_{\vec{k}_\alpha}}}
(a_{\vec{k}_\alpha}+a^\dagger_{-\vec{k}_\alpha})
\delta_{\sum \vec{k}_i,0}
\, .
\end{eqnarray}
Here $m$ denotes the mass, 
$\Omega$ the spatial volume, $g$ the coupling constant,
and $\epsilon_{\vec k}$ the single-particle energy
\begin{equation}
\epsilon_{\vec{k}}=\sqrt{\vec{k}^2+m^2}
\quad.
\end{equation}
The field $\phi$ and its conjugate momentum $\pi$ 
are time-independent (Schr\"odinger picture) operators which satisfy the
commutation relations for bosons, and $a$, $a^\dagger$ are the
associated annihilation and creation operators.
While the kinetic part of the Hamiltonian is normal ordered
($:\ldots :$), the interaction is not.
 
When expressed in terms of annihilation and creation
operators 
the Hamiltonian takes on a form which is very similar to that of the
quantum gas Hamiltonian (\ref{quantum_gas}).
More precisely, the $\phi^4$ Hamiltonian {\em contains}
a Bose gas Hamiltonian with two-body interaction matrix element
\begin{equation}
\langle lk|V|ji\rangle_+
=\left({4\atop 2}\right)
{g\over 4!\Omega\sqrt{\epsilon_i\epsilon_j\epsilon_k\epsilon_l}}
\delta_{\vec{k}_i+\vec{k}_j,\vec{k}_k+\vec{k}_l}
\quad.
\end{equation}
The derivation of a flow
equation for $g$ can now proceed in the same vein as that for $U$,
again starting from Eq. (\ref{mod_bosons}).
Now, however, apart from $2\to2$ particle scattering the
$\phi^4$ Hamiltonian with its novel interactions
$a^\dagger a^\dagger a^\dagger a$, $a^\dagger a^\dagger a^\dagger a^\dagger$
etc.
also permits $2\to4$ and $2\to6$ scattering.
Therefore, in Eq. (\ref{mod_bosons}) the intermediate state
may be not just $|ab\rangle$, but also
$|abik\rangle$, $|abil\rangle$, $|abjk\rangle$, $|abjl\rangle$
or $|abijkl\rangle$.
As long as the magnitude of the external momenta is negligible
compared to the cutoff, it is in all six cases $\epsilon_a=
\epsilon_b=\epsilon_\Lambda$ and 
\begin{equation}
\langle lk|V|\ldots\rangle_+ \langle\ldots |V|ji\rangle_+ =
{g\over 4\Omega\epsilon_\Lambda^2}
\langle lk|V|ji\rangle_+ \delta_{\vec{k}_b,-\vec{k}_a}
\,\, .
\end{equation}
Hence in order to account for the larger set of allowed intermediate states
we merely have to
introduce an extra factor $6$, and obtain thus
\begin{equation}
\Delta g= 
-{d\ln\epsilon_\Lambda \over d\ln\Lambda}\,
{3\rho(\epsilon_\Lambda)\over 8\Omega\epsilon_\Lambda^2}\,
g^2\cdot \Delta s
\quad.
\end{equation}
For $\Lambda\gg m$ it is
$\epsilon_\Lambda=\Lambda$, 
$\rho(\epsilon_\Lambda)=\Omega\epsilon_\Lambda^2/2\pi^2$, and the
flow equation reduces to
\begin{equation}
\Delta g= -{3g^2\over 16\pi^2}\Delta s
\quad,
\end{equation}
in agreement with the well-known 1-loop result for the
$\beta$ function of $\phi^4$ theory.\cite{fisher,generalize}
\subsection{Screening of fermion-fermion interactions}
We consider nonrelativistic fermions in spatial
dimension $d$ ($d\ge 2$) which interact through a two-body interaction
\begin{eqnarray}
\langle lk|V|ji \rangle_- 
&=& 
[V(\vec{q})\delta_{s_l s_i}\delta_{s_k s_j}
- V(\vec{q}')\delta_{s_k s_i}\delta_{s_l s_j}]
\nonumber \\
&&
\,\times
\delta_{\vec{K}_i+\vec{K}_j,\vec{K}_k+\vec{K}_l}
\quad,
\label{screen}
\end{eqnarray}
duly antisymmetrized to account for Fermi statistics,
and with $\{s_\alpha\}$ denoting the spin quantum numbers
and $\vec{q}, \vec{q}'$ 
the respective momentum transfers
\begin{eqnarray}
\vec{q}&:=&\vec{K}_l-\vec{K}_i=\vec{K}_j-\vec{K}_k
\quad,
\nonumber \\
\vec{q}'&:=&\vec{K}_k-\vec{K}_i=\vec{K}_j-\vec{K}_l
\quad.
\end{eqnarray}
We investigate scattering processes for which
\begin{equation}
0 < |\vec{q}|,\Lambda \ll |\vec{q}'|, |\vec{K}_i+\vec{K}_j|, K_F
\quad.
\end{equation}
In this regime only the ZS contribution (\ref{zs}) can
significantly modify the two-body interaction;
its cross term ZS' (Eq. (\ref{zs_prime})),
as well as the BCS contribution (\ref{bcs}),
are suppressed by a factor $\Lambda/K_F$.
This can be seen directly from the geometry of the 
Fermi surface.
The three constraints on the intermediate state:
(i) both $\vec{K}_a$ and $\vec{K}_b$ 
lie in the cutoff shell of thickness $2\Lambda$;
(ii) more stringently, one of them lie in the infinitesimal shell to
be eliminated; and 
(iii) $\vec{K}_a-\vec{K}_b=-\vec{q}'$ (ZS') or
$\vec{K}_a+\vec{K}_b=\vec{K}_i+\vec{K}_j$ (BCS), respectively ---
reduce the momentum space volume available to the internal momentum
$\vec{K}_a$ to $O(K_F^{d-2}\Lambda\Delta\Lambda)$.
In contrast, for $|\vec{q}|\sim \Lambda$ 
the ZS contribution
with its condition $\vec{K}_a-\vec{K}_b=-\vec{q}$ 
allows a momentum space volume
of the order $K_F^{d-1}\Delta\Lambda$.

To evaluate the ZS contribution 
at some given momentum transfer $\vec{q}$,
we first  define the angle $\vartheta$ between $-\vec{q}$ and
the internal momentum $\vec{K}_a$,
\begin{equation}
\cos\vartheta\equiv z:=-{\vec{q}\cdot \vec{K}_a \over |\vec{q}||\vec{K}_a|}
\quad,
\end{equation}
change coordinates from original ($\vec{K}$) to little ($\vec{k}$) momenta,
and write, up to corrections of order $|\vec{q}|/K_F$,
\begin{eqnarray}
\epsilon_a-\epsilon_b = v_F (|\vec{K}_a|-|\vec{K}_b|)
&=& v_F (\sigma_a |\vec{k}_a| -\sigma_b |\vec{k}_b|)
\nonumber \\
&=& v_F |\vec{q}| z
\end{eqnarray}
with $v_F$ denoting the Fermi velocity.
Next we note that the term with $\theta(\sigma_a)\theta(-\sigma_b)$
and the term with $\theta(-\sigma_a)\theta(\sigma_b)$ yield identical
contributions; therefore it suffices to consider only
the first term and then multiply it by two.
Finally, assuming that in the interaction matrix element (\ref{screen})
the cross term is negligible,
\begin{equation}
|V(\vec{q}')| \ll
|V(\vec{q})|
\quad,
\end{equation}
the two matrix elements in Eq. (\ref{zs}) can simply be replaced
by $V(\vec{q})^2$ modulo Kronecker symbols for spin and momentum
conservation.
By virtue of these Kronecker symbols one
of the two summations over internal modes collapses trivially,
leaving
\begin{eqnarray}
\Delta V(\vec{q})
&=& 
-{2\over v_F} \Delta\left[
\sum_a
{\theta(\Lambda-|\vec{k}_a|)\theta(|\vec{k}_a|-|\vec{q}|z+\Lambda)}
\right.
\nonumber \\
&&
\left.
\quad\quad\quad
\times
{{\theta(\sigma_a)\theta(|\vec{q}|z -|\vec{k}_a|)}
\over |\vec{q}|z}
\right]
\cdot V(\vec{q})^2
\quad.
\end{eqnarray}
Here the first two $\theta$ functions explicitly enforce
the sharp cutoff constraint for both $\vec{k}_a$ and $\vec{k}_b$
($|\vec{k}_a|,|\vec{k}_b|\le\Lambda$),
while the latter two
$\theta$ functions enforce $\sigma_a=1$
and $\sigma_b=-1$, respectively.
Under these constraints it is always $\vartheta\in [0,\pi/2)$
and hence $z>0$.

The above equation can be immediately integrated from cutoff
$\Lambda\gg |\vec{q}|,|\vec{k}_a|$ (symbolically,
$\Lambda\to\infty$)
down to
$\Lambda\ll |\vec{q}|,|\vec{k}_a|$
(symbolically, $\Lambda\to 0$),
to yield the total modification of the two-body interaction:
\begin{eqnarray}
\lefteqn{{1\over V_{\rm eff}(\vec{q})}
-{1\over V_{\rm bare}(\vec{q})}
=}
\nonumber \\
&&
{2\over v_F}\!
\left.
\sum_a
{\theta(\Lambda-|\vec{k}_a|)\theta(|\vec{k}_a|-|\vec{q}|z+\Lambda)
\theta(\sigma_a)\theta(|\vec{q}|z -|\vec{k}_a|)
\over |\vec{q}|z}
\right|_{0}^{\infty}
\nonumber \\
&& \mbox{}
\end{eqnarray}
At the lower bound ($\Lambda\to 0$) the various conditions imposed by the
$\theta$ functions cannot all be satisfied simultaneously, and therefore
the product of $\theta$ functions vanishes.
At the upper bound ($\Lambda\to\infty$), 
on the other hand, the cutoff constraints
imposed by the first two $\theta$ functions are trivially
satisfied and thus can be omitted.
In this case the sum over $a$ is evaluated by turning it 
into two integrals, one over a radial
variable such as $|\vec{k}_a|$ or $\epsilon_a$, the other
over the solid angle.
At a given solid angle and hence given $z$, the fourth
$\theta$ function restricts the radial
integration to the range $|\vec{k}_a|\in [0,|\vec{q}|z]$ or,
equivalently,
$\epsilon_a\in [0,v_F |\vec{q}|z]$.
This energy interval in turn corresponds to a number
$[\rho(\epsilon_F) v_F |\vec{q}| z]$ of states,
$\rho(\epsilon_F)$ being the density of states
at the Fermi surface.
(It includes a factor to account for the spin degeneracy.)
The integration over the solid angle
is constrained to a semi-sphere, due to $\vartheta\in[0,\pi/2)$,
and hence reduced by a factor $1/2$ as compared to a full-sphere
integration.
Altogether we obtain
\begin{equation}
{1\over V_{\rm eff}(\vec{q})}
-{1\over V_{\rm bare}(\vec{q})}
=
{2\over v_F}
\,\, {1\over2}\,\,
\rho(\epsilon_F) v_F |\vec{q}|z\,
{1\over |\vec{q}|z}
=
\rho(\epsilon_F)
\end{equation}
and thus
\begin{equation}
{V_{\rm eff}(\vec{q})}=
\left[
{1\over V_{\rm bare}(\vec{q})}+ \rho(\epsilon_F)
\right]^{-1}
\quad.
\end{equation}
This result describes the well-known screening of
fermion-fermion interactions.\cite{fetter}
\subsection{BCS instability}
Our last example pertains to fermions with an attractive pairing
interaction
\begin{equation}
\langle lk|V|ji \rangle_- = -V\, \delta_{\vec{K}_j,-\vec{K}_i}
\delta_{\vec{K}_l,-\vec{K}_k}
[\delta_{s_l s_i}\delta_{s_k s_j} - \delta_{s_k s_i}\delta_{s_l s_j}]
\, ,
\end{equation}
which is the simplest form of BCS theory.\cite{schrieffer}
Due to the pairing condition $\vec{K}_j=-\vec{K}_i$,
$\vec{K}_l=-\vec{K}_k$
it is impossible to satisfy
in the ZS and ZS' terms the requirement that at least one
of the internal modes be in the eliminated shell.
Hence only the BCS term (\ref{bcs}) can modify
the coupling constant.
In the BCS term there are contributions with $\theta(\sigma_a)\theta(\sigma_b)$
and $\theta(-\sigma_a)\theta(-\sigma_b)$, respectively, which yield identical
results; therefore it suffices to consider only
the first contribution and then multiply it by two.
The pairing condition implies $\epsilon_a=\epsilon_b=\epsilon_\Lambda$,
which for modes in the upper ($\sigma=+1$)
eliminated shell is given by $\epsilon_\Lambda=v_F \Lambda$.
The eliminated shell itself
covers an infinitesimal energy interval of width
$[v_F \Lambda\Delta s]$,
which in turn corresponds to a number
$[\rho(\epsilon_F) v_F \Lambda\Delta s]$ of states.
Of the two summations over internal modes one 
collapses trivially due to momentum and spin conservation,
leaving
\begin{eqnarray}
\Delta V
&=&
{V^2\over 2v_F\Lambda}
\Delta \left[\sum_{a} {\theta(\sigma_a)}\right]
\nonumber \\
&=&
{V^2\over 2v_F\Lambda}\,
\rho(\epsilon_F)\, v_F \Lambda\Delta s
= {\rho(\epsilon_F)\over 2}\,V^2 \cdot \Delta s
\quad.
\end{eqnarray}
 
From this flow equation for the BCS coupling $V(s)$
we immediately conclude that as long as the initial coupling $V(0)$ is positive,
$V(s)$ diverges as $s\to\infty$.
This indicates the occurence of binding (``Cooper pairs'')
at very low temperatures.
Furthermore, we can again convince ourselves that
the sequence of effective theories retains complete information
about the system's low-energy dynamics:
low-energy observables
such as the zero-temperature gap\cite{schrieffer}
\begin{equation}
\Delta_0 =
2\Lambda(s) \exp\left[
-{2\over \rho(\epsilon_F) V(s)}\right]
\end{equation}
do not depend on the flow parameter $s$
and are thus unaffected by the successive mode elimination.
\section{Rescaling}\label{renormalization}
So far we have only considered the systematic elimination of modes.
Yet there is a second operation, namely rescaling, 
which in the context of renormalization
plays an equally important role.
In this section we shall first discuss the rescaling operation by itself,
and later its interplay with the elimination of modes.

Again we would like to couch the rescaling procedure 
into the language of linear operators in Hilbert space.
To this end we define a unitary
dilatation operator $D(s)$ which,
if acting on a (box-normalized) boson excitation
(\ref{bose_excite}) or fermion excitation (\ref{fermi_excite}),
simply rescales all momenta by $e^{s}$:\cite{isometric}
\begin{equation}
D(s)|\vec{k}_1^{(\pm)}\ldots \vec{k}_n^{(\pm)}\rangle
:= |e^{s}\vec{k}_1^{(\pm)}\ldots e^{s}\vec{k}_n^{(\pm)}\rangle
\quad\forall\, 
s\ge 0
\, .
\end{equation}
This operation can be viewed in two different ways, analogous to
the active and passive interpretations of coordinate
transformations:
either as an enlargement of the physical momenta at fixed length unit
(active interpretation), or as an enlargement of the length unit at
fixed physical momenta (passive interpretation).
The latter view might be pictured as the observer with a television camera
``zooming out'' away from the probe, thereby watching on the camera's monitor
a larger section of the probe and excitations with
apparently smaller wavelengths. 

Let the state of a physical system 
be described by a canonical statistical operator 
\begin{equation}
\rho(b)\propto \exp[-H(\tilde{\Omega})/b]
\quad,
\end{equation}
where we have written explicitly the possible dependence of 
the Hamiltonian on a finite volume $\tilde{\Omega}$.
After ``zooming out'' the canonical distribution will appear
distorted, namely as
\begin{equation}
\rho_{\rm zoom}(b,s) := D(s) \rho(b) D^\dagger(s)
\quad.
\end{equation}
We wish to cast this zoomed distribution again into a
canonical form,
\begin{equation}
\rho_{\rm zoom}(b,s)\propto 
\exp[-H_{\rm zoom}(e^{-d\cdot s}\tilde{\Omega},s)/b(s)]
\quad,
\end{equation}
with a modified external parameter $b(s)$ and
some modified, ``zoomed'' Hamiltonian $H_{\rm zoom}$.
Since the rescaling makes the system appear smaller on the observer's (imaginary)
monitor, the modified Hamiltonian must now depend on the rescaled
volume $e^{-d\cdot s}\tilde{\Omega}$ ($d$ being the spatial dimension)
rather than on $\tilde{\Omega}$.
The thus defined zoomed Hamiltonian is easily determined:
\begin{equation}
H_{\rm zoom}(\Omega,s)=
{b(s)\over b}\,D(s) H(e^{+d\cdot s}\Omega) D^\dagger(s)
\quad.
\end{equation}
In general, the $s$-dependence of the external parameter $b$ is not
a priori known and must be fixed by additional constraints.
One such constraint might be the requirement that some
given reference Hamiltonian $H^{\rm ref}$ be scale invariant,
$H^{\rm ref}_{\rm zoom}=H^{\rm ref}$.

Just like the effective Hamiltonian obtained via mode elimination,
the zoomed Hamiltonian generally contains modified masses, chemical potential and
coupling constants; however, it does not contain any qualitatively new interactions.
How the various parameters evolve as the flow parameter $s$ increases,
is again governed by a set of differential equations,
the so-called scaling laws.

To illustrate these general ideas we re-consider our first example,
bosons with point interaction.
Making use of the identity
\begin{equation}
D(s) a^{(\dagger)}(\vec{k}) D^\dagger(s) = a^{(\dagger)}(e^{s}\vec{k})
\end{equation}
for annihilation and creation operators,
we find for the free part of the Hamiltonian
\begin{equation}
H^{(0)}_{\rm zoom}(\Omega,s)=
\sum_{\vec{k}} \left[ {b(s)\over b}\epsilon(e^{-s}\vec{k})\right]
\, a^\dagger(\vec{k}) a(\vec{k})
\end{equation}
and for the interaction
\begin{equation}
V_{\rm zoom}(\Omega,s)=
\left[ {e^{-d\cdot s} b(s)\over b}\right]\, V(\Omega)
\quad.
\end{equation}
Taking $b=k_B T$,
assuming a relativistic single-particle energy
\begin{equation}
\epsilon(\vec{k})=\sqrt{ \vec{k}^2 +m^2} - \mu
\end{equation}
with mass $m$ and chemical potential $\mu$,
and requiring that in the limit $m=\mu=0$ the free dynamics become
scale invariant ($H^{(0)}_{\rm zoom}=H^{(0)}$), we obtain the scaling relations
\begin{equation}
T(s)=e^s T
\quad,\quad
m(s)=e^s m
\quad,\quad
\mu(s) = e^s \mu
\end{equation}
which in turn imply
\begin{equation}
U(s) = e^{-(d-1)s} U
\quad.
\end{equation}
This last result immediately translates into a scaling law
for the coupling constant $g$ of $\phi^4$ theory. The coupling $g$
differs from $U$ essentially by a factor 
$\sqrt{\epsilon_i\epsilon_j\epsilon_k\epsilon_l}$, which 
just leads to an
extra factor $e^{2s}$ in the scaling:
\begin{equation}
g(s) = e^{(4-(d+1))s} g
\quad.
\end{equation}
As expected, the number $\epsilon:=4-(d+1)$ in the exponent is the deviation
of the space-time dimension from the upper critical dimension 
four.\cite{fisher}

The combination of rescaling and mode elimination constitutes a complete
renormalization group transformation.
In the by now familiar spirit we define the renormalized
statistical operator
\begin{equation}
\rho_{\rm ren}(b,s):=
D(s) P(s) \rho(b) P(s) D^\dagger(s)
\quad,
\end{equation}
which is obtained by first eliminating 
high momentum modes in the shell $[e^{-s}\Lambda,\Lambda]$
and then rescaling all momenta so as
to recover the original cutoff.
In our intuitive picture this transformation corresponds to
first coarse-graining the physical spatial resolution
and then ``zooming out,'' thereby rescaling the length unit such that
the {\em apparent} spatial resolution, defined in rescaled rather
than original length units, stays constant.
(One might visualize the apparent spatial resolution as the
fixed pixel size of the
observer's monitor, which, as the camera zooms out,
corresponds to ever worse physical resolutions.)
As in the previous cases
we wish to cast the renormalized statistical operator into
the canonical form
\begin{equation}
\rho_{\rm ren}(b,s) \propto
\exp[-H_{\rm ren}(e^{-d\cdot s}\tilde{\Omega},b(s),s)/b(s)]
\end{equation}
with modified external parameter $b(s)$ and
renormalized Hamiltonian $H_{\rm ren}$.
How the various parameters in the renormalized
Hamiltonian (masses, couplings, etc.)
evolve under the combined action of mode elimination and rescaling,
is governed by the full renormalization group equations.
These can be obtained by adding the respective
flows due to mode elimination and rescaling.

The rescaling part of the full renormalization group equations 
is specifically adapted to the study
of critical phenomena.
Critical phenomena are distinguished by the lack of an
intrinsic length scale:
as the observer is ``zooming out,'' at fixed apparent resolution,
he keeps seeing 
the long-wavelength dynamics governed by the same 
masses and coupling constants.
In other words, the observer has no means of inferring
from these parameters the scale at which
the dynamics is taking place.
Critical phenomena are therefore naturally associated with 
a fixed point --the ``critical'' fixed point-- of
the full renormalization group equations.
 
If one studies effects 
{\em other} than critical phenomena,
such as screening or the BCS instability, then
there is generally no reason for a repeated
rescaling.
\section{Conclusion}
In this paper we have formulated the renormalization program
exclusively in terms of algebraic operations in Hilbert space.
This new formulation has given us the opportunity to 
expose the close relationship between Wilson's renormalization
and the projection technique \`a la Bloch and Feshbach, and to consider
some of the ideas underlying renormalization from
a slightly different conceptual perspective.
Aside from these formal developments we 
have also explored the potential for practical uses of the
algebraic approach, and found that in four examples 
--bosons with point interaction, $\phi^4$ theory, 
screening of fermion-fermion
interactions, and BCS instability--
the flow equations for the respective couplings
could be derived in an efficient and straightforward manner.

The algebraic approach carries the potential for interesting
future developments.
First, it is easily generalized 
since
projection operators permit the elimination not just of
short-wavelength modes, but also of
other kinds of information deemed irrelevant,
such as high angular momenta, spin degrees of freedom,
or entire particle species.
Secondly, the algebraic approach builds a bridge 
between renormalization and
macroscopic transport theories, as also the transition
from micro- to macrodynamics can be effected by means of
very similar projection techniques.\cite{rau}
It may thus provide a natural language for the
study of issues such as the renormalization of
macroscopic transport
theories, or effective kinetic theories.\cite{jeon:yaffe}
\acknowledgements
I thank A. Beli\'c, H. Forkel, A. Molinari, B. Mottelson and K. Sailer for
helpful discussions.
Financial support 
by the European Union HCM fellowship programme is gratefully
acknowledged.


\begin{references}
\bibitem{trace}
The reduced statistical operator is not a statistical operator
in the conventional sense, since
generally $\mbox{tr}\rho_{\rm eff}\ne 1$.
\bibitem{info_loss}
Discarding these irrelevant pieces of the density matrix generally entails
a loss of information as to the system's microstate.
The amount of information lost can be quantified by comparing 
the entropies before and after coarse graining.
\bibitem{loops}
In conventional renormalization the Taylor expansion in powers
of $b$ corresponds to a loop expansion.
Our truncation of
this expansion immediately after the first term thus amounts to
a 1-loop approximation.
\bibitem{generalize}
A more detailed derivation of the renormalization group equations
for $\phi^4$ theory by means of projection techniques,
as well as the generalization to arbitrary
spatial dimension, can be found in Ref. \CITE{muller:rau}.
\bibitem{isometric}
At finite volume (whence a discrete spectrum) or in the presence of
a momentum cutoff the dilatation operator should be defined with more care,
as some of the rescaled momenta $\{e^s \vec{k}_i\}$ might no longer be
admissible in the given Hilbert space.
\end{references}
\end{document}